\documentclass[nohyperref]{article}

\usepackage{microtype}
\usepackage{graphicx}
\usepackage{subfigure}
\usepackage{booktabs} 

\usepackage{hyperref}

\usepackage[accepted]{icml2022}

\usepackage{amsmath}
\usepackage{amssymb}
\usepackage{mathtools}
\usepackage{amsthm}

\usepackage[capitalize,noabbrev]{cleveref}

\theoremstyle{plain}

\theoremstyle{definition}

\theoremstyle{remark}

\DeclareMathOperator*{\argmax}{arg\,max}

\usepackage[textsize=tiny]{todonotes}

\icmltitlerunning{Reconstructing gene expression and knockout effect from mutation (Mut2Ex)}

\begin{document}

\twocolumn[
\icmltitle{Reconstructing gene expression and knockout effect scores from DNA mutation (Mut2Ex): methodology and application to cancer prediction problems}



\icmlsetsymbol{equal}{*}

\begin{icmlauthorlist}
\icmlauthor{$\text{Maya Ramchandran}^{*}$}{comp}
\icmlauthor{Maayan Baron}{comp}
\end{icmlauthorlist}

\icmlaffiliation{comp}{ZephyrAI, Virginia, USA}

\icmlcorrespondingauthor{Maya Ramchandran}{maya@zephyrai.bio}

\icmlkeywords{Machine Learning, Partial Least Squares, Cancer prediction}

\vskip 0.3in
]



\printAffiliationsAndNotice{}  

\begin{abstract}
Building prediction models for outcomes of clinical relevance when only a limited number of mutational features are available causes considerable challenges due to the sparseness and low-dimensionality of the data. In this article, we present a method to augment the predictive power of these features by leveraging multi-modal associative relationships between an individual's mutational profile and their corresponding gene expression or knockout effect profiles. We can thus reconstruct expression or effect scores for genes of interest from the available mutation features and then use this reconstructed representation directly to model and predict clinical outcomes. We show that our method produces significant improvements in predictive accuracy compared to models utilizing only the raw mutational data, and results in conclusions comparable to those obtained using real expression or effect profiles.  
\end{abstract}

\section{Introduction}
\label{intro}
\par Utilizing somatic mutation data to predict clinically relevant patient outcomes can yield suboptimal results due to the sparse information content in binary hotspot mutation features \cite{prasad2016perspective, west2016no}. Results are considerably worse in scenarios where the number of genes assayed is small, as is the case with patient data derived from commercially available NGS panels. \cite{shen2015clinical}. However, the vast availability of such data alongside clinical outcomes measured on the same patients could be enormously useful in developing tools to improve diagnoses and treatment decisions, and therefore necessitates solutions to augment the granularity of mutation data in order to most effectively train prediction models for such outcomes \cite{hodis2012landscape, cancer2012comprehensive, van2002gene}. We show that great improvements in this direction can be made by first leveraging the correlation between hotspot mutation status and the gene expression profile of a given sample to construct a latent representation of expression space directly from mutation. 
\par It has been previously established across many biological applications that higher level genomic features such as gene expression typically have higher discrimination and predictive power for phenotypes and other clinical features than lower level features such as somatic mutation, regardless of the machine learning model applied \cite{costello2014community, menden2019community, chiu2019predicting}. We demonstrate that a more effective utilization of mutation data first reconstructs expression profiles by exploiting the biological relationship between the two modalities, and then directly uses this reconstructed representation to train prediction models for any outcome of interest such as patient survival, cancer subtyping, and disease staging. We additionally show that it is possible to reconstruct scores measuring the effect of gene knockout directly from mutation using a similar framework. 
\par In order to accomplish this, we propose a modeling approach (\textit{Mut2Ex}) based on partial least squares regression, a popular statistical framework that models the common structure shared by the dependent variables and predictors in the presence of potential multi-collinearity between features \cite{wold1985}. To our knowledge, \textit{Mut2Ex} is the first method to transform binary mutation data into a continuous, data-rich representation directly based on gene expression. Unlike other continuous-valued embeddings of binary samples, ours is comparable with the output of gene expression assays and can be used in conjunction with or instead of patient gene expression data, either within model construction or evaluation. The intuition behind our approach is drawn from the biological connection between the mutation status of a gene and its expression, as well as the interplay between genes within pathways and other correlative relationships. To that end, \textit{Mut2Ex} is capable of jointly inferring the expression state of all genes of interest as opposed to requiring separate prediction models for each gene’s expression; this allows shared information to be borrowed across genes in order to increase overall efficiency and accuracy. 
\begin{figure}[t]
\vskip 0.2in
\begin{center}
\centerline{\includegraphics[width=.8\columnwidth]{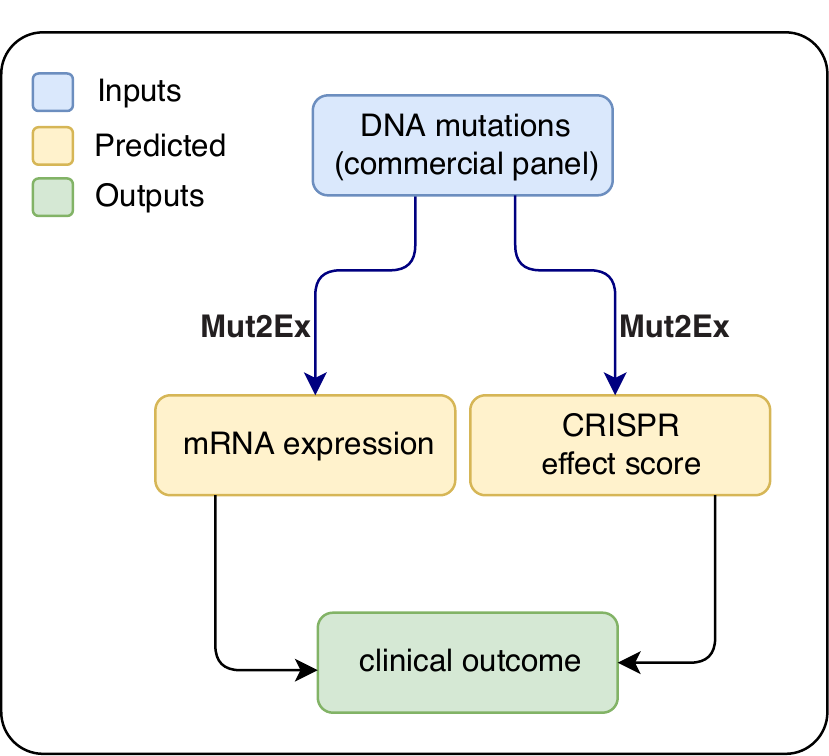}}
\caption{\textbf{\textit{Mut2Ex} workflow}: binary DNA mutation status is used to reconstruct either mRNA expression or CRISPR knockout effect scores using \textit{Mut2Ex}. The reconstructed output can then be used to train models predicting various clinical outcomes.}
\label{fig1-diagram}
\end{center}
\vskip -0.2in
\end{figure}
\par Previous work in this area focuses on the use of partial least squares to directly handle high dimensional gene expression data in inferential and prediction contexts, particularly in the case in which there are significantly more features than samples \cite{boulesteix2006, nguyen2004, yang2017}. Liquet et. al. describe an approach to study the relationship between different types of high-dimensional 'omics data or between 'omics data and phenotypes \yrcite{liquet2015}. However, their focus is on the integration of multiple available modalities for the same set of samples and on the grouping of features in a high-dimensional context, as opposed to our aim of inferring the expression profile of a sample given only the mutation status of a relatively small number of genes. Other work in modeling joint relationships between mutation and expression typically involves the use of deep learning models that not only require a large amount of training samples, computational complexity, resources, and time, but also ultimately lack interpretability \cite{avsec2021, zhou2015}. In contrast, our proposed approach is fast, efficient, relatively simple, and interpretable. Our model can be trained using as few as a couple hundred samples, and can handle input feature sets either larger or smaller than the sample size without sacrificing accuracy. 
\par We begin this article by describing the \textit{Mut2Ex} methodology and the contexts for which it is designed. We additionally highlight the biological and mathematical intuition involved in the formulation of the approach. We then present applications for which our method improves upon the current standard, particularly in the prediction of clinically relevant outcomes from low-dimensional commercial mutation panels measured on cancer patients. We show that our framework can be used to reconstruct either gene expression or CRISPR knockout effect scores \cite{behan2019}, and result in similar conclusions for the ultimate prediction task of interest to what would be obtained if we had access to the true expression or effect profiles. For all applications, we show that our approach is far superior to using the mutation profiles directly for the same prediction problems. We note that while we have chosen to focus our methodological description on reconstructing gene expression for clarity, the same concepts transfer when reconstructing gene knockout effect scores. 

\section{Methods}
\subsection{Notation}
Let $\mathbf{X} \in \mathbb{R}_{n \times p}$ and $\mathbf{Z} \in \mathbb{R}_{n \times q}$ be two data matrices containing $n$ observations (rows) of $p$ predictors (mutation) and $q$ variables (gene expression), respectively. For both $\mathbf{X}$ and $\mathbf{Z}$, each predictor and variable represents a gene; these gene sets do not necessarily need to be overlapping. Now, let $\mathbf{X}^{*} \in \mathbb{R}_{m \times p}$ be a data matrix containing $m$ observations of the same $p$ predictors as $\mathbf{X}$ (mutation) for which we have a corresponding vector $\mathbf{y} \in \mathbb{R}_{m \times 1}$ containing the clinical outcome to be modeled. We denote by the subscript $c$ the centered form of a matrix; that is, for matrix $M \in \mathbb{R}_{n \times p}$, $M_c = M - \frac{1}{n} \mathbf{e}\mathbf{e}^T M$, where $\mathbf{e} \in \mathbb{R}_{n \times 1}$ is an n-length vector of $1$'s. 
\subsection{Partial Least Squares Regression}
Partial least squares (PLS) techniques are increasingly popular in genomic applications, primarily because they have been designed to handle the situation in which there are far more (potentially correlated) features than samples. A significant advantage of partial least squares is its explicit focus on capturing the joint correlation between the input and output features in an efficient latent representation suited both for prediction tasks and dimension reduction; this is particularly effective at integrating multiple 'omics feature sets measured on the same samples. 
Partial least squares regression (PLSR) is based on the following latent component decompositions of the centered predictor and response matrices: 
\begin{align}
    \mathbf{X_c} &= LP^T + E \\
    \mathbf{Z_c} &= LQ^T + F
\end{align}
where $L \in \mathbb{R}_{n \times c}$ is the matrix of latent components for a given number of components $c$, with the columns representing each component's scores across all n training observations. $P \in \mathbb{R}_{p \times c}$ and $Q \in \mathbb{R}_{q \times c}$ are matrices of coefficients for $\mathbf{X_c}$ and $\mathbf{Z_c}$, respectively, and $E \in \mathbb{R}_{n \times c}$ and $F \in \mathbb{R}_{n \times q}$ are the corresponding matrices of residuals. 
\par The foundation of PLSR is in modeling the latent components matrix $L$ as a linear combination of $\mathbf{X_c}$; that is, 
\begin{align}
    L = \mathbf{X_c}W
\end{align}
where $W \in \mathbb{R}_{p \times c}$ is a matrix of weights; optimizing $W$ is the primary objective in PLSR algorithms. Once $W$ is determined, the latent component matrix $L$ is then used to predict $\mathbf{Z_c}$ in place of the original variables within $\mathbf{X_c}$, where $Q^T$ is the least squares solution of equation (1); that is,
\begin{align}
    Q^T &= \left(L^TL\right)^{-1}L^T\mathbf{Z_c}
\end{align}
Now, incorporating equations (3) and (4) into (1), we can express the regression equation relating $\mathbf{X_c}$ to $\mathbf{Z_c}$ as 
\begin{align}
    \mathbf{Z_c} &= \mathbf{X_c}W\left(L^TL\right)^{-1}L^T\mathbf{Z_c} + F \nonumber \\
    &= \mathbf{X_c}B + F
\end{align}
where the matrix $B = WQ^T = W\left(L^TL\right)^{-1}L^T\mathbf{Z_c} \in \mathbb{R}_{p \times q}$ contains the PLS regression coefficients. The corresponding fitted response matrix $\mathbf{\hat{Z}}$ is written as
\begin{align}
    \mathbf{\hat{Z}} &= \mathbf{X_c}B \nonumber \\
    &= L\left(L^TL\right)^{-1}L^T\mathbf{Z_c}
\end{align}
representing the least squares solution of a linear regression predicting $\mathbf{Z_c}$ from $L$. Finally, to obtain predictions $\mathbf{\widehat{Z^*}} \in \mathbb{R}_{m \times q}$ given a new a matrix of new, uncentered observations $\mathbf{X^*} \in \mathbb{R}_{m \times p}$, we compute
\begin{align}
    \mathbf{\widehat{Z^*}} &= \left(\mathbf{X^*} - \frac{1}{n} \mathbf{e}
    \mathbf{e}^T \mathbf{X_c}\right)B + \frac{1}{n} \mathbf{e}\mathbf{e}^T \mathbf{Z_c}
\end{align}
where $\mathbf{e}$ again represents the n-length vector of $1$'s. 
\par Given this formulation, it is clear that the specification of $L$, and thus $W$, determines the basis of all components required to produce PLSR predictions of a gene expression matrix $\mathbf{Z}$ from a mutation matrix $\mathbf{X}$. The basic idea in PLS-based approaches that the latent components $L$ are designed to have a high covariance with the response $\mathbf{Z}$. This is clear in examining the objective function for optimizing W: 
\begin{align}
    W_i &= \argmax_w w^T\mathbf{X_c}^T\mathbf{Z_c}\mathbf{Z_c}^T\mathbf{X_c}w \nonumber \\
    &= \argmax_w \sum_{j = 1}^q \text{Cov}^2\left(\mathbf{Z_c}_j, \mathbf{X_c}w \right)
\end{align}
for $i = 1, \hdots, c$, with $W_i$ and $\mathbf{Z_c}_j$ representing the i$^{th}$ j$^{th}$ columns of $W$ and $\mathbf{Z_c}$, respectively. Therefore, $W$ is constructed by finding the linear combination of the input features that have maximal squared covariance with each dimension of the response. Popular algorithms to solve for $W$ with appropriate constraints in the multivariate response context include NIPALS and SIMPLS; in our applications, we use an implementation of SIMPLS, although we do not anticipate major changes in performance between algorithms \cite{wold1975path, dejong1993}. 

\subsection{Reconstructing gene expression profiles}
The PLSR framework described above is advantageously designed in handling the biological relationships between genes both within and across modalities. The optimization of the weights matrix $W$ takes into account the correlative relationships between the gene expression values of each output gene across samples with the corresponding mutation status of the input genes. Additionally, the correlation between genes both within each mutational profile and within each gene expression profile are jointly handled through the optimization of the linear combination coefficients within $W$: each resulting linear combination of the features within $\mathbf{X}$ (thus capturing between-gene correlation) is explicitly designed to have maximal covariance with each feature (gene) in $\mathbf{Z}$. The ability of PLSR to handle the potential singularity of $\mathbf{X}^T \mathbf{X}$ when $p > n$ or the existence of multicollinearity between the features in $\mathbf{X}$ allows for flexibility in specifying the input mutational gene set, regardless of the number of samples $n$ available for training. 
Given the overall structure and benefits to PLSR in this context, we can thus build a model $z(\mathbf{X^*})$ to reconstruct the expression profiles $\mathbf{\widehat{Z^*}}$ of a set of samples given their mutational profiles $\mathbf{X^*}$ as follows: 
\begin{align}
     z\left(\mathbf{X^*} \right) = &\bigg[\left(\mathbf{X^*} - \frac{1}{n} \mathbf{e}
    \mathbf{e}^T \mathbf{X_c}\right)W\left(W^T \mathbf{X_c}^T\mathbf{X_c}W\right)^{-1} \nonumber \\
    &\times \left(W^T \mathbf{X_c}^T\mathbf{Z_c}\right)\bigg] + \frac{1}{n} \mathbf{e}\mathbf{e}^T \mathbf{Z_c}
\end{align}
once $W$ has been solved as in equation (8). We refer to model $z\left(\mathbf{X^*} \right)$ as \textit{Mut2Ex}.  

\subsection{Predicting clinical outcomes}
Now, we can build a prediction model for the clinical outcome $\mathbf{y}$ directly from the reconstructed expression $\mathbf{\widehat{Z^*}}$ (the output of \textit{Mut2Ex}) as opposed to the original mutation $\mathbf{X^*}$. That is, if we denote by $y = f(\cdot)$  any regression function relating an input feature set to $\mathbf{y}$ (for example, regularized regression or machine learning approaches such as Random Forest, Gradient Boosted Trees, or Neural Networks), we can train a model
\begin{align}
    \mathbf{\hat{y}} &= f\left(\widehat{Z^*}\right)  \nonumber \\
    &= f\left(z(\mathbf{X^*})\right)
\end{align}
instead of using the raw mutation features directly (i.e. $\mathbf{\hat{y}} = f\left(\mathbf{X^*}\right)$). As we show in the data application, the increase in granularity afforded by the continuously-valued reconstructed expression over sparse binary mutation data greatly improves the prediction performance of most regression approaches.

\subsubsection{Choosing the number of components}
\par The number of components $c$ with which to construct $L$ plays a critical role in determining a PLSR model. The maximal number of latent components that can have non-zero covariance with $\mathbf{Z}$ is $c_{\text{max}} = \min(n-1, p)$; for genomic data in which often $p > n$, setting $c = c_{\text{max}}$ results in a fully saturated, and thus overfit, model. The the number of components $c$ must therefore be chosen to strike a balance between generalizability and preserving variability in the reconstructed expression profile, since the lower the number of components, the more the predictions regress to the mean. Additionally, if this reconstructed expression is also intended to be used within a subsequent model to predict the clinical outcome $\mathbf{y}$, it is critical that the degree of variability in the reconstruction be appropriately aligned with the variability in $\mathbf{y}$. When the ultimate task is estimating $\mathbf{y}$, it is less important that $\mathbf{\widehat{Z^*}} = z(\mathbf{X^*})$ perfectly capture the true expression profile $\mathbf{Z^*}$ in absolute value than that it reflect the correlative relationships between genes in $\mathbf{Z^*}$ that are most predictive of $\mathbf{y}$. In this case, we suggest using a cross-validation approach to choosing $c$ by minimizing the distance (using an appropriate metric) between $\mathbf{y}$ and $f\left(z(\mathbf{X^*})\right)$ as opposed to $z(\mathbf{X})$ and $\mathbf{Z}$. For the results we present in Section 3, we found that restricting the model to between 40 and 50 components produced the best overall prediction outcomes; however, we emphasize that the choice of this hyperparameter is context specific and should be tuned to the particular application of interest. 

\subsubsection{Feature set determination}
The choice of features for which to reconstruct gene expression is dependent on the ultimate prediction task and the gene set available within the input mutation data. We have designed \textit{Mut2Ex} to effectively handle relatively low-dimensional mutational feature sets corresponding to the size of most commercial panels, and as previously mentioned, the genes to be reconstructed in expression space do not need to necessarily correspond with the available genes in mutation space. To determine an effective reconstructed expression gene set to predict clinical outcome $\mathbf{y}$ typically requires a combination of biological knowledge (such as previously identified driver genes) and data-driven variable selection procedures. In general, since we are using a far more sparse genomic representation of the samples to reconstruct continuous gene expression values, more effective PLSR predictions are obtained for $q < p$; that is, utilizing more features in mutation space to reconstruct fewer features in expression space. 

\section{Results}

\subsubsection{Training setup}
To apply the \textit{Mut2Ex} method to biological data, we selected a training setup based on mutation and expression data measured on 691 cell lines from the Cancer Cell Line Encyclopedia \cite{barretina2012cancer}. We hypothesized that learning the relationships between the two modalities in the cell line context would capture the biologically meaningful correlations to be extrapolated without potential additional noise present in tumor data which could lead to overfitting and lack of generalizability. When subsequently applied to real tumor mutation data, the reconstructed expression would then be more likely to contain the relevant biological variation required for further inference or prediction tasks that would have been present in the true expression profiles were they measured, even if the two sets are not identical in value. We posited this translation would be effective since the  mutational profiles inputted into the model tend to be far more stable between cell lines and tumor samples than gene expression. However, we underscore that \textit{Mut2Ex} may be trained on pre-clinical or real tumor 'omics data depending on the context, and that our particular choice of cell line training data was designed for the applications we have chosen to present. Finally, we note that our implementation of \textit{Mut2Ex} utilizes functions provided by the {\tt PLS} package in R \cite{plspackage}. 

\subsection{Breast cancer classification}
\subsubsection{PAM50 gene expression signatures}
\par To investigate the performance of \textit{Mut2Ex} on clinical data, we examined the use of reconstructed expression in classifying breast cancer tumors into the PAM50 classification subtypes. PAM50 describes clinically meaningful intrinsic molecular subtypes defined by the mRNA expression of 50 key genes, and has been shown to significantly improve predictions of prognosis compared to other genomic signatures or tumor characteristics \cite{nielsen2010comparison, parker2009supervised, filipits2014}. This classification is widely applied clinically to categorize breast cancer tumors into the following 5 subtypes: Luminal A, Luminal B, human epidermal growth factor receptor 2 (HER2)-enriched, Basal-like, and Normal-like. As the subtypes are based on gene expression signatures of the established PAM50 genes, classification of a new tumor sample typically requires assaying the expression level of these same genes using RNA-sequencing, microarray, or qPCR and comparing the similarity of the resulting expression profile to the signatures defined by each subtype. 

\begin{figure}[t]
\vskip 0.2in
\begin{center}
\centerline{\includegraphics[width= \columnwidth]{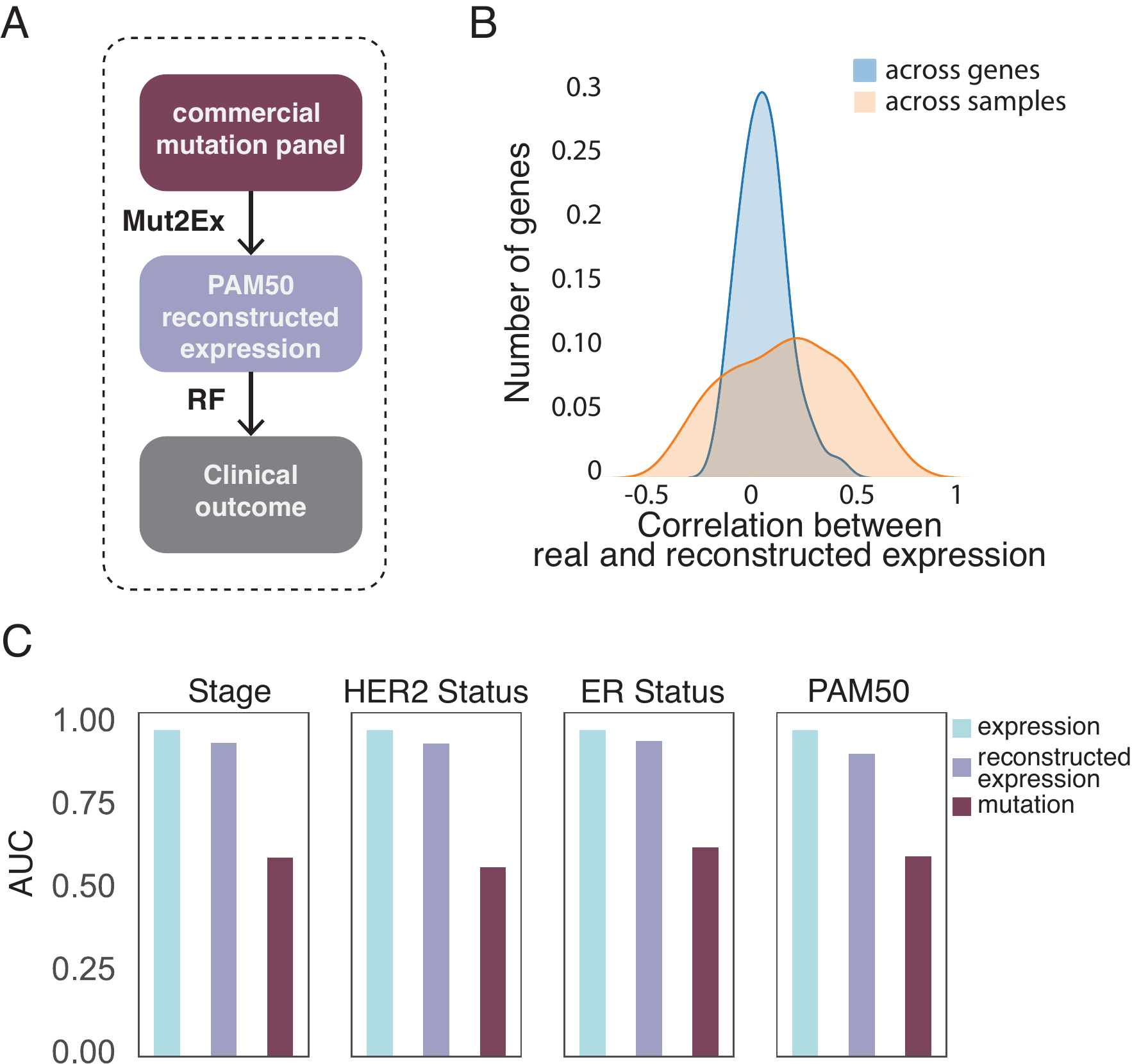}}
\caption{\textbf{Mut2Ex application in various BRCA clinical classification tasks}: (A) Illustration of the ML workflow: \textit{Mut2Ex} was applied to the TCGA-BRCA mutation set subsetted to the genes on the Foundation One panel in order to reconstruct the expression of the PAM50 signature genes. The output was then used to build individual models to predict various clinical outcomes. (B) Histogram of Pearson correlation coefficients between real and reconstructed expression of all 50 PAM50 genes, across genes (blue) \& across BRCA tumor samples (orange). (C) Area under the ROC-curve (AUC) for classification models trained on either expression (teal), reconstructed expression (blue), or mutation (red).}
\label{fig2-brca}
\end{center}
\vskip -0.2in
\end{figure}

\subsubsection{Evaluating \textit{Mut2Ex}-based clinical subtyping performance}

\par The clinical importance of PAM50 underlies the highly beneficial impact of being able accurately classify tumors into the PAM50 subtypes based solely on the commercial mutation panels most commonly measured in practice. To achieve this, we applied \textit{Mut2Ex} on The Cancer Genome Atlas Breast Cancer (TCGA-BRCA) project data, a compendium containing whole genome sequencing and RNA-Seq data as well as predicted PAM-50 subtypes (based on the relevant RNA-seq signatures) for 571 total breast cancer patient samples \cite{cancer2012comprehensive}. We subsetted the mutation features to the 324 genes measured on the FoundationOne CDx diagnostic panel and used \textit{Mut2Ex} to reconstruct the expression of the 50 genes comprising the PAM50 signature (Figure \ref{fig2-brca}A) \cite{milbury2022clinical}. Only 11 of the PAM50 genes are included in the FoundationOne panel, so we were primarily reconstructing expression for genes for which we did not have the corresponding mutation statuses. We then trained a Random Forest (RF) classifier on the reconstructed expression to predict the PAM-50 molecular subtypes and compare these predictions with the assigned subtypes provided by TCGA-BRCA. We additionally considered the tasks of predicting tumor stage, HER2 status, and ER status. Finally, we note substituting other common diagnostic panels such as MSK-IMPACT and DFCI-ONCOPANEL within the \textit{Mut2Ex} workflow in place of FoundationOne led to similar conclusions across all prediction problems. 

\par For all objectives, we compared the prediction accuracy of RF learners trained on \textit{Mut2Ex}-produced reconstructed expression to RF learners trained on either the true expression profiles for the PAM50 genes provided by TCGA-BRCA or the original binary mutation profiles for the FoundationOne panel genes (Figure \ref{fig2-brca}C). Across the board, the models built on reconstructed expression perform very similarly to models built on true expression and generate highly accurate predictions (AUCs $> .9$), whereas models built on mutation features perform far worse (AUCs $< .6$); these results are particularly remarkable given that outcomes such as PAM50 classification are explicitly based on expression. We emphasize the concordance in prediction performance despite the relatively low correlation between the true and reconstructed expression profiles (average pearson correlation of 0.05 and 0.17 across genes and samples, respectively, Figure \ref{fig2-brca}B) even for specific genes known to be important in breast cancer pathology. Nevertheless, \textit{Mut2Ex} is still clearly able to extract the meaningful information from mutation status relevant for clinical outcome prediction by leveraging the associative relationships between mutation and expression. 

\subsection{Gene knockout effect prediction}
\subsubsection{Application to gene dependency scores}
\par Although precision medicine has been traditionally considered to be a direct derivative of genomics, the majority of patients do not harbor actionable mutations as currently defined. Moreover, when genomic data from patient tumors is clinically actionable, most patients who get treated solely according to their genomic panel results in practice do not benefit significantly overall and require additional measurement of phenotypic features to receive impactful clinical interventions \cite{letai2017functional}. A far more informative--albeit impractically clinically obtainable--metric is a measure of how dependent a tumor cell is on a gene for survival, also known as a dependency score, which can be measured by genetically knocking out a gene in cancer cell models using the CRISPR/Cas-9 system and comparing the proliferation of those models to their unperturbed controls. Applied genome-wide and to a large cohort of cancer cell lines, these functional genetic perturbation screens have already led to the identification of important cancer oncogenes and tumor suppressor genes, revealed the broad essentiality of some genes to cell fitness and the context specificity of others, and helped map the association of genes to functionally distinct and highly biologically relevant pathways \cite{aguirre2016genomic, barbie2009systematic, tsherniak2017defining, cowley2014parallel, marcotte2012essential, marcotte2016functional}.

\par The dependency score represents an interpretable metric of perturbation response and can itself be considered an analyzable outcome in treatment determination. There is consequently a need to improve gene knockout effect score prediction models and for such models to be simple and explainable to inform high stakes decisions \cite{rudin2019stop}. To this end, we adapted the \textit{Mut2Ex} framework to reconstruct gene knockout effect scores directly from commercial mutation panels, given the biological correlation between mutation and effect we expect to see similarly to expression. Unlike the application of \textit{Mut2Ex} in reconstructing expression to predict further variables, in this case we considered the reconstruction of effect scores as the ultimate objective. As in our breast cancer analysis, we restricted the mutation features for training \textit{Mut2Ex} to the FoundationOne diagnostic panel in order to jointly reconstruct knockout effect scores for 78 genes that have been shown to be involved in tumorigenesis and tumor invasion; there are 53 genes in common between the two sets. Since the knockout experiments underlying the scores were all performed on cell lines, we limited our reconstruction to the CCLE cell line universe (n = 939) using a cross-validation framework so as to not overlap training and test lines. 

\begin{figure}[t]
\vskip 0.2in
\begin{center}
\centerline{\includegraphics[width= \columnwidth]{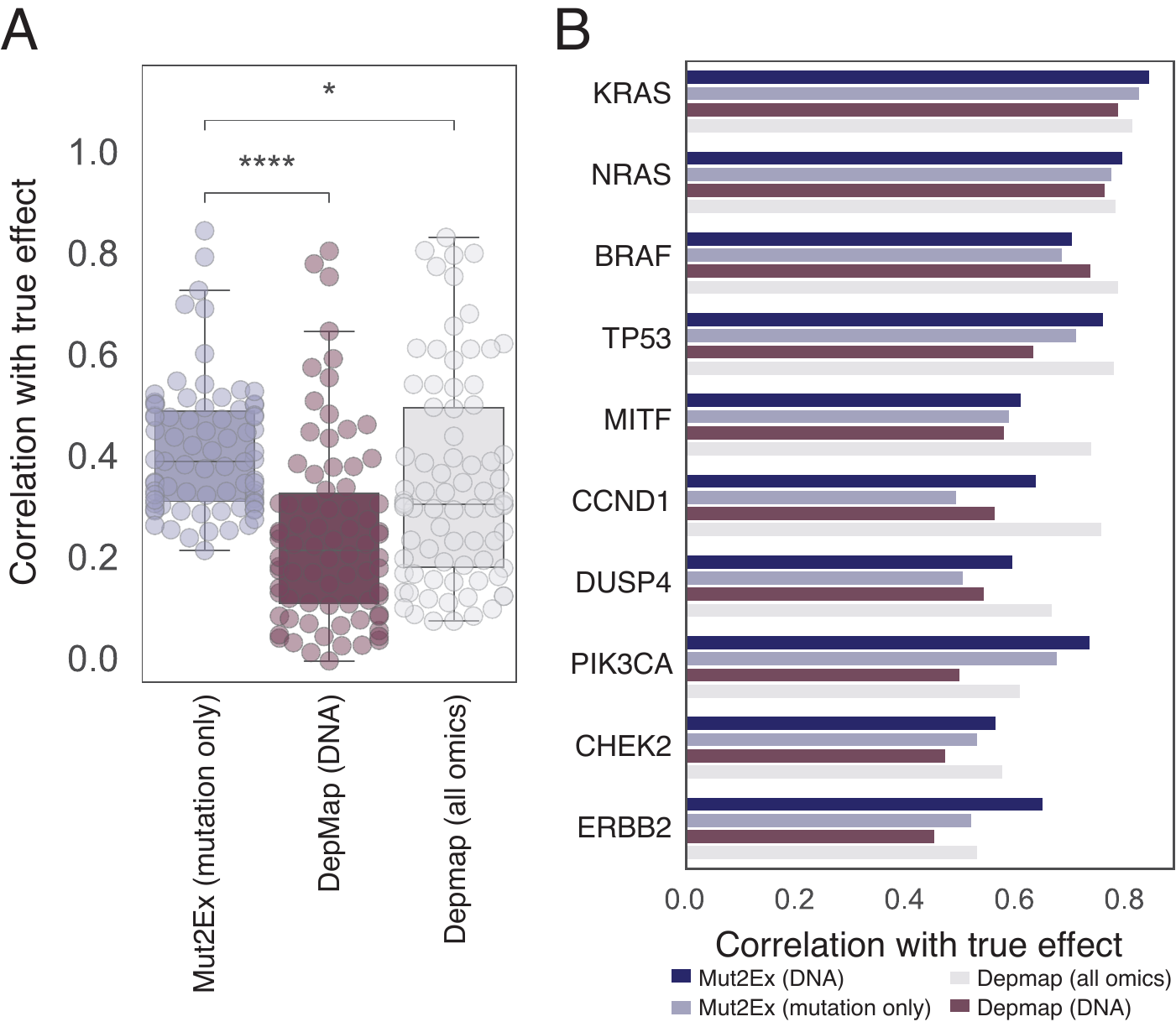}}
\caption{\textbf{Predicting gene knockout effect scores using \textit{Mut2Ex}}: (A) Box plot of Pearson correlation coefficients between reconstructed and true knockout effect scores of the selected 78 genes using Mut2Ex (light purple), DepMap using DNA features only (dark purple) or DepMap using all features (light gray). Significance was determined by a two sample t-test (**** P$< 10^{-4}$) (B) Bar plots of the Pearson correlation coefficients of the top 10 performing genes by DepMap DNA feature-model.
}
\label{fig3-effect}
\end{center}
\vskip -0.2in
\end{figure}

\subsubsection{Comparison to modeling approach by DepMap}

The Dependency Map (DepMap) project from which the effect scores were obtained additionally provides performance metrics for models they trained on various types of 'omics features to predict dependency scores across nearly 20,000 genes. We accordingly chose to compare the correlation between the reconstructed scores from \textit{Mut2Ex} and the true effect scores with the correlations achieved by the DepMap 'omics models. 

\par DepMap considered a total of 181,951 features in training their models, including mutation status, RNA-Seq, Copy Number Variation (CNV), methylation profiling, gene fusion, and tissue annotation \cite{Dempster2020}. They investigated different training paradigms given their various feature sets; in particular, an `all omics' model drawing from every available feature type listed above, and a `DNA-only' model for which only DNA-based features (such as mutation and CNV) were given as inputs. For all approaches, feature selection was first performed by training a regularized regression model individually for each gene knockout to identify the top 1000 correlated features. A Random Forest (RF) model was then trained on the selected features to predict the dependency scores for the given gene across all cell lines. DepMap reported that other machine learning models such as elastic net produced similar results to RF. By comparing the performance across their 'omics models and identifying the top correlated features for each perturbation, DepMap concluded that RNA-Seq features are by far the most predictive of gene effect. We note that unlike \textit{Mut2Ex}, DepMap's approach in training separate models for each gene knockout did not leverage any relationships between gene dependencies. 

\subsubsection{Evaluating \textit{Mut2Ex}-based dependency score reconstruction}

\par For our analysis, we employed two \textit{Mut2Ex} models; the first was trained only on mutation features as previously described, whereas the second additionally included CNV features for the same set of panel genes in order to examine the potential additional signal carried by copy number alterations within this context. Similarly to previous studies, we used the Pearson correlation coefficient between the reconstructed and true effect scores as our performance metric \cite{Dempster2020, ben2020predicting}. All CNV data was categorized into 3 classes representing deletions, neutral variations, and amplifications. Figure \ref{fig3-effect} displays the results of our two \textit{Mut2Ex} models and how they compare to DepMap's models in predicting the chosen 78 genes. We find that the reconstructed effect scores from the most parsimonious mutation-only \textit{Mut2Ex} model significantly outperform all DepMap models, including those trained on RNA-seq features in the `all-omics' paradigm (Figure \ref{fig3-effect}A). Focusing on the top performing genes from the `DNA-only' DepMap models (Figure \ref{fig3-effect}B), we observe that \textit{Mut2Ex} still produces more accurate predictions than DepMap in almost all cases. For the majority of genes, the addition of CNV features to \textit{Mut2Ex} improves prediction performance over just mutation, indicating the benefit of including these features in training when available; however, even the mutation-only \textit{Mut2Ex} model (trained on 323 features as opposed to the nearly 200k considered in DepMap's largest models) still predominantly surpasses the current standard. Overall, these results show the flexibility and applicability of \textit{Mut2Ex} to extend the use of DNA molecular features in precision medicine. 

\section{Discussion}
\par In this article, we have demonstrated the ability of \textit{Mut2Ex} to significantly improve the predictive signal gleaned from limited numbers of mutational features across a variety of clinical applications. By using a PLS-based framework to construct a continuous, expression (or effect)-based representation from mutational features, \textit{Mut2Ex} explicitly takes advantage of the two primary types of correlation we expect to see from a biological standpoint; first, correlations between genes within each modality, and secondly, correlations between gene sets across modalities. Jointly reconstructing genes of interest in expression space allows \textit{Mut2Ex} to efficiently borrow shared information across genes using a relatively small number of samples, resulting in better prediction performance than methods that handle genes separately or require large sample sizes to train deep learning models. In predicting gene knockout effect scores, we show the improvements afforded by this joint reconstruction over DepMap's separate prediction models for each gene while training on the same set of cell lines as \textit{Mut2Ex}.

\par Other considerable advantages of \textit{Mut2Ex} are the interpretability of the reconstruction model and flexibility in managing various available input features. Interpretability is a critical factor in deploying machine learning models to make clinical decisions in practice, and thus was a priority in developing \textit{Mut2Ex}'s ability to handle commercial mutation panels. Previous characterizations interpreting the coefficients and latent components from PLS-based models can be applied to \textit{Mut2Ex} to determine the mutation features most important in reconstructing expression or effect \cite{kvalheim2010interpretation, chun2010sparse, tran2014interpretation, boulesteix2006}. We leave formalizing the explanatory aspects of \textit{Mut2Ex} as a direction for future research. Furthermore, other available 'omics features such as copy number variation can be included within training regardless of whether they measure the same genes as in the mutation set. Our results established that the addition of CNV features boosts reconstruction accuracy for key genes and often produces equivalent or better accuracy than models evaluating hundreds of thousands of features (including gene expression). However, we emphasize that these types of additional features are not necessary for strong model performance. 

\par We have presented two different applications of \textit{Mut2Ex} to either reconstruct gene expression or knockout effect scores. For the former, the low degree of correlation between reconstructed and true expression was ultimately unimportant in light of the high overall prediction accuracy for the clinical variables of interest. For the latter, we considered the predicted effect scores as the final output, and could reliably do so since the reconstructed effect from \textit{Mut2Ex} had significantly higher correlation with the true effect scores than in the expression case. We posit this pattern is largely due to the nature of the two types of measurements rather than the degree of overlap between the input mutation features and the genes to be reconstructed, as neither the effect score nor the expression correlation distributions shift for genes included versus excluded from the FoundationOne panel; further exploration of this assertion is an intended future research direction. Gene expression is biologically an intermediary in influencing phenotypes (hence why regression models trained on expression tend to be superior in predicting phenotypic variables), whereas the knockout effect scores represent a measure of overall cell viability in response to perturbations and therefore themselves capture phenotypic changes. In addition, gene expression measurements are subject to variability depending on factors such as equipment choice, batch, and time of measurement in relation to cellular processes, resulting in distributional shifts across studies \cite{luo2010comparison, lazar2013batch}. The stability of knockout effect scores in relation to gene expression gives compelling support as to why \textit{Mut2Ex} is able to generalize the learned correlation between mutation and effect in absolute value more so than for expression. Even though distributional differences between the training and test expression data result in lower correlation values, the biologically meaningful relationships between mutation and expression in relation to phenotypic variables are preserved. 

\par Because \textit{Mut2Ex} exhibits this characteristic, an additional benefit to training prediction models on reconstructed in place of true expression is the potential for greater generalizability to new datasets. This paradigm requires that every test dataset first be processed by \textit{Mut2Ex} to construct a representation of expression compatible with the prediction model, thus conceivably mitigating dataset shift and batch effects. Since \textit{Mut2Ex} takes mutation as its input (a modality far less variable than expression), building regression models on reconstructed expression may reduce confounding influences and standardize training and test data in relation to a unified set of correlative structures learned by \textit{Mut2Ex}. Investigating this premise further is an area for future research.

\section{Conclusion}

We have presented a method for transforming clinical panels measuring mutational features to a continuous representation based on gene expression or knockout effect. Conventional gene panels are not typically informative enough to drive clinical decisions on their own; however, we show that through \textit{Mut2Ex}, such panels carry enough signal to generate reconstructed expression (or effect) representations that are capable of producing clinically actionable results. Furthermore, this representation is independent of the available mutation features, in the sense that the mutation features may be disjoint from the representation features and specifically selected to optimize the subsequent prediction task. We have also detailed the mathematical formulation of \textit{Mut2Ex} and the intuition behind how a PLSR-based approach effectively leverages underlying biological correlations between genes in mutation and expression space even in a low-dimensional setting. Through our two data applications, we have illustrated the considerable benefit afforded by reconstructed expression and effect produced by \textit{Mut2Ex} in prediction tasks of clinical importance compared to models based solely on mutation. Overall, we have shown \textit{Mut2Ex} to be a powerful tool in more efficiently utilizing the vast amount of existing clinical panel data to improve treatment outcomes and derive actionable insights. 
Future directions include devising a standard method to optimize the reconstruction gene set based on the available mutation features, as well as identifying an appropriate metric to quantify the application-specific predictive ability of the reconstructed expression for hyperparameter tuning. 

\section*{Acknowledgements}
We would like to thank Emily Vucic, Dillon Tracy, and Jeff Sherman for their insightful comments, as well as Felicia Kuperwaser and Yoni Schoenberg for their contributions to the data analysis.


\bibliography{example_paper}
\bibliographystyle{icml2022}


\end{document}